\documentclass{aa}

\usepackage{natbib}
\usepackage{longtable}
\usepackage{lscape}
\usepackage{amsmath}
\usepackage{txfonts}
\usepackage{color}
\usepackage{url}
\usepackage{xspace}

\usepackage{graphicx}
\bibpunct{(}{)}{;}{a}{}{,}

\newcommand{\corot}{\textsl{CoRoT}\xspace}

\begin{document}

   \title{Detecting transits from Earth-sized planets around Sun-like stars}

   \author{S. Carpano
          \inst{1}
          \and
	  M. Fridlund
          \inst{2}}

   \offprints{S. Carpano, e-mail: scarpano@sciops.esa.int}
   \institute{XMM-Newton Science Operations Centre, ESAC, ESA, PO Box 50727, 28080 Madrid, Spain
             \and
	     ESA Astrophysics Missions Division, Research and Scientific Support Department, ESTEC, PO Box 299, 2200 AG, Noordwijk, The Netherlands}

   \date{Submitted: 18 December 2007; Accepted: 25 March 2008}
   
   \abstract{Detecting regular dips in the light curve of a star is an easy way to detect the presence of an
   orbiting planet. \corot is a Franco-European mission launched at the end of 2006, and one of its main objectives
   is to detect planetary systems using the transit method.}{In this paper, we present a new method for transit
   detection and determine the smallest detected planetary radius, assuming a parent star like the Sun. }{We
   simulated light curves with Poisson noise and stellar variability, for which data from the VIRGO/PMO6 instrument 
   on board SoHO were used. Transits were simulated using the Universal Transit Modeller software. Light curves were
   denoised by the mean of a low-pass and a high-pass filter. The detection of periodic transits works on light curves
   folded at several trial periods with the particularity that no rebinning is performed after the folding. The
   best fit was obtain when all transits are overlayed, i.e when the data are folded at the right period.}{Assuming
   a single data set lasting 150\,d, transits from a planet with a radius down to 2\,R$_\oplus$ can be
   detected. The efficiency depends neither  on the transit duration
   nor on the number of transits observed. Furthermore we simulated transits with periods close to 150\,d in 
   data sets containing three observations of 150\,d, separated by regular gaps
   with the  same length. Again, planets with a radius down to 2\,R$_\oplus$ can be
   detected.}{Within the given range of parameters, the detection efficiency depends slightly on the apparent magnitude of 
   the star but neither on the transit duration nor
   the number of transits. Furthermore, multiple observations might represent a solution for the \corot mission for 
   detecting small planets when the
   orbital period is much longer than the duration of a single observation.}

     \keywords{Stars: planetary systems  -- Occultations  -- Techniques: photometric -- Methods: data analysis } 

\maketitle
%

\section{Introduction}
\label{sec:int}
 Beginning with the discovery by \cite{Mayor1995} 12 years ago of a giant planet, 51 Peg $b$, around its Sun-like host star,
the search for planets outside the solar system has become a major topic in astrophysics. 
Among the various methods available to search for exoplanets, the transit method presents the opportunity of directly determining 
the planet's radius relative to that of its parent star and, if more than 
one transit is observed, its orbital period. It also has the advantage of beeing possible to monitor many thousands 
of target stars simultaneously. This is of course necessary, since the probability that a planet transits its star depends 
on the radii of star and planet, as well as the orbital distance and the eccentricity \citep{Rauer2007}. It is never a 
large number, {\it viz.}

\begin{equation}
\text{Probability} = 0.0045 \left({1AU \over a}\right) \times \left({r_s + r_p} \over {1_\odot}\right) \times 
\left[{1 + e~cos\left({\omega} \over 2\right)} \over {1 - e^2}\right],
\end{equation}
where $a$ is the semi major axis of the planetary orbit, $r_s$ and $r_p$ are the stellar and planetary radii respectively, $e$ the
eccentricity, and $\omega$ the longitude of the periastron. 
 It is easy to see that, for an Earth analogue, the probability of a transit beeing observed is about 0.5\%, while for a similar planet at 0.05~AU, 
 the geometrical probability rises to about 10\%. When also folding in the poorly known factor 
 $\eta_{Planet}$~(number of planets per star system), one realizes this need for (simultaneous) observations of large numbers of stars.
 
Diminishing of the light curve relative to the light from the star, the photometric signal $P$, is of course immediately translatable into 
the ratio of planetary to stellar radius
 
 \begin{equation}
 P = {{\Delta S} \over {S}} = \left[{R_\text{Planet} \over R_\text{Star}}\right]^2.
 \end{equation}
 Therefore, the minimum size of the planet that can be detected will only depend on the precision with which one can
 measure the photometric output from the star and know the stellar radius.
 
 Successful results have been obtained for a score of transiting objects, since the first 
 detection by \cite{Char00}, but all of these are essentially Jupiter-like objects in very close orbits 
 (i.e. less than a few days).  If we want to detect and study planets more akin to our own Earth, we are 
 facing formidable problems, where the stellar radius currently can only be discerned from model calculations, 
 and even then  is still contains severe uncertainties. The achievable photometric precision is governed by the problem of 
 acquiring long and un-interupted sequences of data. When observing from  the ground we are faced with enormous 
 challenges. In contrast, by launching a telescope into space, into an orbit allowing these long,
  uninterrupted data acquisitions,  much higher precision can be achieved. Nevertheless, by improving 
  the photometric signal, one also increases the ambition. Thus, with the advent of space-based assets, 
  the goal is now set at the actual detection of a planet with the same radii as our own Earth.  
 
 Simultaneously, during the development of the first space missions, a number of transit-detection algorithms 
 have been proposed in the literature including Bayesian algorithms 
\citep[i.e,][]{Aigrain2002}, matched filters \citep[i.e,][]{Jenkins1996}, box-shape transit finder 
\citep{Aigrain2004}, and the box-fitting least squares method \citep{Kovacs2002}. A comparison of the capabilities 
of existing methods was first made by \cite{Tingley2003a} (revised in Tingley~2003b) and later by \cite{Moutou2005} on updated versions of 
the algorithms. Besides the detection method, filtering of photon noise and stellar variability also play an 
important role. In \cite{Carpano2003}, we used an optimal filter to significantly reduce the stellar 
variability and enhance the S/N ratio of the transits. However, this filter uses a 
reference signal and hence some a priori information of the shape of the signal we are looking for.

 The \corot spacecraft was launched on December 27, 2006 and has been carrying out routine scientific operations since February 2, 2007.  
 \corot stands for COnvection, ROtation ,and planetary Transits. Briefly, this satellite is carrying out ultra high-precision 
 photometry for the dual purpose of studying asteroseismological signals from stars and both discovering and studying extra-solar 
 planets with unprecedented precision. The focal plane is divided into two segments, one each for these objectives. Each 
 of the parts of the focal planet (designated `exo' and `seismo') is equipped with two CCD-based detectors. The exoplanetary 
 detector records the light curve of up to 12\,000 targets. Of these, 20\% have a time sampling of 32\,sec, while the 
 rest are sampled every 512\,sec. Continuous monitoring of up to 150\,days for each selected field is carried out, and a 
 grism in front of the detector allows thousands of the light curves to be obtained in three colours for stars with visual 
 magnitudes from about 12 to 16. For further basic information about the mission we refer to ESA SP-1306 (2006, The \corot Mission) 
 for a description of all aspects as of pre-launch.

 One of the main goals of \corot  is to detect smaller exo-planets than previously possible with any method. 
 Radial velocity measurements is currently limited to the detection 
 of planets 5-10 times as massive as the Earth itself in orbits around very low-mass stars \citep{Udry2007}. 
 The occultation method as carried out from the ground also has limitations imposed by both the atmosphere and, 
 more important, by the interruptions inherent in any lightcurve obtained from the surface, although extensive networks developed in
recent years alleviate this. In this case, the 
 depth in the light curve that can be obtained from the ground under the best of circumstances is at best a 
 few tenths of a percent \citep{Rauer2007}, representative of planets of either the Jupiter or Saturn class. In the 
 case of \corot, on the other hand, it has been shown by in-orbit verification that the data is photon-noise-limited, 
 or almost so, in the whole magnitude range for the exo-planetary segment for integration times 
 shorter than 1 hour \citep{Baglin2007}.  The detections of Earth-sized planets then becomes possible with 
 the \corot mission -- at least for shorter orbital periods -- but is nevertheless is a challenging task.

In this paper we present a new, although simple, transit detection method that could be applied to \corot data 
after having applied a smoothing and detreding filter. 
We do not attempt to compare the efficiency of the method with respect to other  ones, for which a proper analysis should be performed much
like the one in  
\cite{Tingley2003a} and \cite{Moutou2005}. Here we concentrate on describing the method, and quote its capabilities and its efficiency
at detecting small planets when few transits are observed. The detection is applied to simulated light curves 
with Sun-like stellar variability, lasting 150\,days as is the case for \corot. Assuming the radius of the parent star is 
fixed to solar value, we check the radius the planet needs for beeing
detected. Stellar apparent magnitude, transit duration and orbital period,  hence the number of transits, 
are all taken to be random in a given range of values. To avoid a long computation time, we limited our simulation to
100 light curves for each stellar activity level and therefore did not follow the approach of \cite{Tingley2003a},
which uses Monte Carlo simulations.

 We also tested the transit detection method in the case of three observations of the 
same object separated by 150\,days, where only one transit is visible in each data set. This simulates the observation 
of the same region at several separated epochs.

The paper is organised as follows. Section~\ref{sec:simul} describes the way the simulated input light curves were derived. In Section~\ref{sec:filter}
we describe the filtering and the transit detection method we have applied. The performances of the method are evaluated in 
Sects.~\ref{sec:results} and~\ref{sec:gaps}, for the single and multiple observations, respectively. Finally, the discussion
of our results is presented in Section~\ref{sec:conc}.


\section{Data simulations}
\label{sec:simul}
\begin{figure}
 \resizebox{\hsize}{!}{\includegraphics{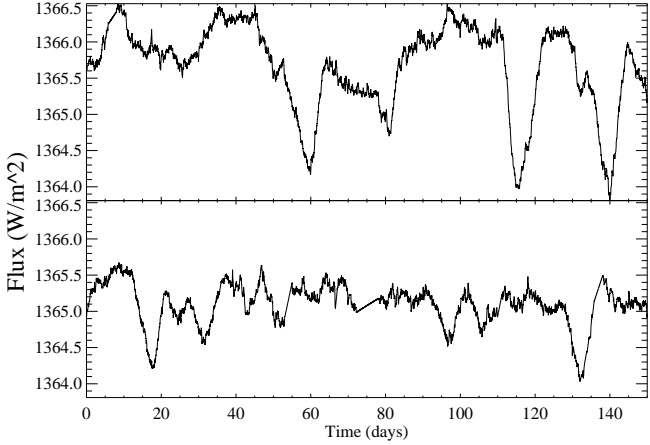}}
\caption{VIRGO/PMO6 data from the SoHO satellite, from year 2000 (top) and 2006 (bottom) used to simulate stellar
activity at high and low level, respectively.}
\label{fig:virgo}
\end{figure}

To test the efficiency of our transit detection method in finding Earth-sized planets  around a Sun-like star,
we simulated light curves using VIRGO/PMO6 data \citep{Froehlich1997} from the SOHO telescope as  done in
\cite{Carpano2003}. Since data from VIRGO are only available on a hourly time grid, we interpolated them to have a data sampling of 10\,min.
The total duration of the light curve is 150\,d as expected from the \corot mission (where the time
resolution is 32\,s or 512\,s). To simulate a high level of stellar activity, VIRGO data from the year 2000 were used, while low 
activity was represented by data from the year 2006. The corresponding VIRGO/PMO6 data sets are shown in Figure~\ref{fig:virgo}.
On the top of these data, we added Poisson noise, assuming that $\Delta F/F=7\times10^{-4}$, where $F$ is the
total number of photons collected in 1\,h for a star of apparent magnitude 15.5. This is what is expected for the \corot mission.
No instrumental noise of any kind has been introduced into the simulated light curves.

\begin{figure}
  \resizebox{\hsize}{!}{\includegraphics{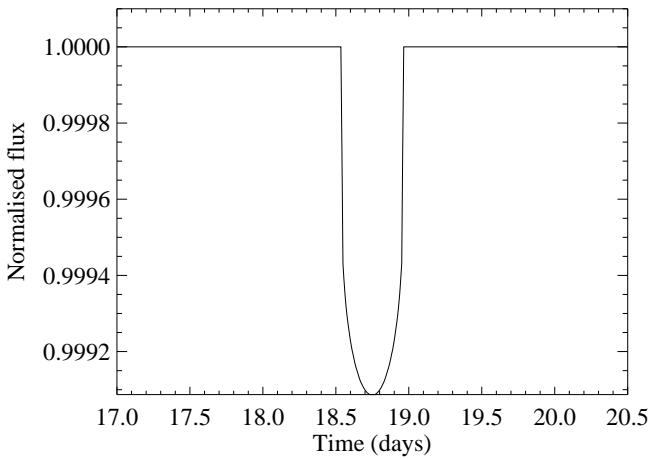}}
 \caption{Output of the UTM code, zoomed around the first transit , where $R_\text{pl}$=3\,R$_\oplus$,
$P$=37.5\,d, $Ph$=0.5,  $tdur$=10\,h. }
 \label{fig:ideal}
\end{figure}

Dips  associated with the transit of a planet are simulated using the Universal Transit
Modeller (UTM) software, developed by \cite{Deeg1999}. The utilised parameters values are listed in Table~\ref{tab:param}:
the mass, $M_\star$, the radius, R$_\star$, and limb-darkening coefficient, $Ld$, of the parent star, which are kept fixed to solar values. The radius of
the planet, $R_\text{pl}$, the orbital period, $P$, and the phase, $ph$, (thus the number of transits), and the transit duration change randomly 
within a given range of values. We require that at least 3 transits are present in the light curves, with a maximum of 6 transits. 
The apparent magnitude, ranging from 12 to 16
is related to the Poisson noise added to the light curve. An example of the output of the UTM code, zoomed around the 
first transit, is shown in Figure~\ref{fig:ideal} where $R_\text{pl}$=3\,R$_\oplus$, $P$=37.5\,d, $Ph$=0.5, $tdur$=10\,h.
The sharp  drop and increase in flux are caused by the ingress/egress of the planet, while the smoother variation close to the bottom is due 
to limb-darkening effects.

\begin{table}
 \centering
 \caption{Parameters used in the Universal Transit Modeller software to simulate the transit of a planet in front of its
 parent star.} 
  \label{tab:param}
 \begin{tabular}{ll}
 \hline  
 Parameter & Value/Range of Value\\
 \hline
 Star mass, $M_\star$ & 1 M$_\odot$ (fixed)\\
 Star radius, $R_\star$ & 1 R$_\odot$ (fixed)\\
 Limb darkening coefficient, $Ld$ & 0.51 (fixed)\\
 Apparent magnitude, $V$ & [12--16]\,mag\\
 Transit duration, $tdur$ & [4.5--36]\,h\\
 Planet radius, $R_\text{pl}$ & [1--5]\,R$_\oplus$\\
 Orbital period, $P$ & [25--50]\,d\\
 Orbital phase, $Ph$ & [0--1]\\
  \hline
   \end{tabular}
\end{table}

\section{Filtering and periodic transit detection}
\label{sec:filter}
\subsection{Filtering}
To filter the Poisson noise and detrend the light curve for stellar variability, we used a low-pass and a high-pass filter, respectively, 
equivalent to a standard box-shaped filter in the Fourier domain. The disadvantage of these filters is that the shape of the dip 
is slightly changing. Without passing 
through the Fourier domain, the data sets are filtered by a boxcar smooth over a certain width, using the 
IDL tool \texttt{smooth}. To reduce the
Poisson noise, we smooth the data over a width of 25 (in units of 10\,min). This width, linked to our data sampling,
limits the lower value of our simulated transit duration (4.5\,h). Outputs from the \corot mission have a data sampling of 32\,s 
(in the oversampled mode), therefore decreasing
 the lower limit for the transit duration to be detected. The smoothing unfortunately reduces the sharpness of
the transit ingress/egress. This shape's characteristic can be used to distinguish dips from an eclipse with respect
to dips from stellar variability;  however, this will have no effect on our detection method as described below.

\begin{figure}
  \resizebox{\hsize}{!}{\includegraphics{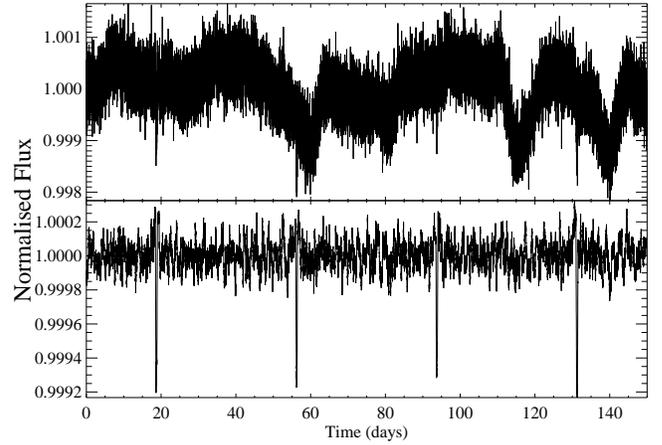}}
 \caption{Simulated light curve before and after filtering. Transits are built with the same parameters as in 
 Figure~\ref{fig:ideal}, stellar variability is taken at the high level, and $V$=12\,mag. }
 \label{fig:filter}
\end{figure}

The detrending is performed in two steps: we first smooth the data over a large width (wider than the maximal transit
duration, i.e. 1.5$\times tdur_\text{max}$=54\,h) and then remove the long-term smoothed light curve to the previous 
short-term smoothed one. This results in a light curve filtered for stellar variability and Poisson noise.
Extreme values are set to the mean light curve value (1 if normalised). 
An example of a light curve before and after filtering is given in Figure~\ref{fig:filter}, where the parameters
for the UTM code are taken as in Figure~\ref{fig:ideal}. Stellar variability is assumed to be high, while the
apparent magnitude V = 12\,mag.

\subsection{Periodic transit detection method}
\label{sec:method}
\begin{figure}
  \resizebox{\hsize}{!}{\includegraphics{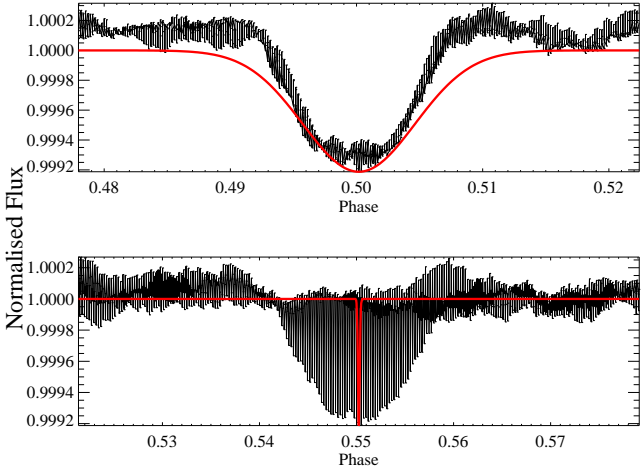}}
 \caption{Light curve folded at the correct (top) and wrong (bottom) period, without any rebinning. Overlayed is
 the best fit Gaussian function (see text for more details).}
 \label{fig:folded}
\end{figure}

Our transit detection method has the advantage of beeing very simple. We first fold the light curve over a given
range of trial periods. The only difference with respect to the conventional epoch-folding method 
\citep{Leahy1983} is that the data are not rebinned after folding. On the folded data set, for each trial period, 
we first sort the data in an ascending order for the phase values and then fit a Gaussian 
function using the IDL function \texttt{gaussfit}. Results are shown in
Figure~\ref{fig:folded}: when data are folded at the correct period, the Gaussian fits the data relatively well
(top of Figure~\ref{fig:folded}). In contrast when data are not folded at the correct period, the width of the
Gaussian function is constrained by the data sampling, which results in a bad fit. 
This is also the case when folding at half of the period, one third of it, etc.
Testing over a full range of
trial periods, the best value is given by the minima of the chi-square function as shown in the next section,  in
Figure~\ref{fig:chitot}.
Note that the fit is performed over the entire folded light curve, while  in Figure~\ref{fig:folded} 
only the region around the transit dip is shown. The shape of the transit is modified by the filter, which 
creates bumps before and after it.

Other minima can be observed at multiple values of the correct period. If the S/N ratio is not too low or if the
period step value is not too high, the chi-square value associated with these sub-harmonics will always be at the secondary minima. 
No signal will be found at half of the period, one third of it, etc., and thus no harmonics are 
present in the periodogram.
The period step should be chosen as small as possible, ideally one time unit, or at least 1/$n$th  of the
transit duration when no more than $n$ transits are expected.

This detection method presents the disadvantage that the position of the fitted Gaussian
function is given by the minima of the filtered light curve. If this minima is not associated with any of the
transits but with some dip coming from the stellar activity, for example, the correct period will not be
found. 

Another property of the method is that the detection efficiency does not depend on the number of transits observed. The right period 
of a data set containing a large number of transits will be detected in the same way that with only a few observed occultations. 
However, the presence of many transits will increase the confidence level of the period detection (see next section).

Comparing with other methods qualitatively, this one has the advantage of not needing to create a set of test light curves to be compared to the
observed one as done for the matched-filter \citep{Jenkins1996} or the method of \cite{Protopapas2005} (which uses analytical function 
to characterise a transit).  It does not need any assumption on any parameter, like the transit duration, as  with the sliding transit 
template correlation developed by Team~1 in \cite{Moutou2005}. 
The non-improvement of period detection with an increase in the number of transits is also observed in the method used by Team~2 in 
\cite{Moutou2005}. They make a box search for all 
data points deviating from the average signal by 3 sigma, remove
spurious events, and subsequently search for a periodicity in all detected 
epochs. Our method has the advantage of doing everything at once. Furthermore, we do not detect harmonics of the periodic signal but only 
the sub-harmonics, which decreases the noise level in the chi-square function. 
Of course a proper comparison would be necessary to test the efficiency of the periodic transit
detection and/or time consumption of this method with respect to existing ones.

\subsection{Confidence levels}

\label{sec:confid}
\begin{figure}
  \resizebox{\hsize}{!}{\includegraphics{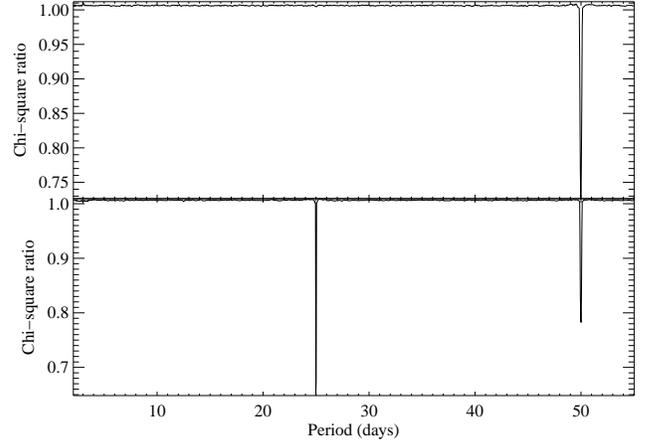}}
 \caption{Chi-square ratio values for the Gaussian fit of light curves folded over trial periods ranging from
 5\,h--55\,d, for an input period at 50\,d (top) and 25\,d (bottom). The minimum of the curve gives the value for the best-fitted period.}
 \label{fig:chitot}
\end{figure}

To determine the confidence level of a period detection, we estimated the ratio
between the  chi-square resulting from a linear fit and the one resulting
from a Gaussian fit.  We call this the $\chi^2_\text{ratio}$ (see Figure~\ref{fig:chitot}, top, for three transits, and bottom, 
for six transits).  The second chi-square increases with the number of 
transits, but the first increases faster. Thus when increasing the number of transits, the $\chi^2_\text{ratio}$ function
goes to lower minima, therefore enhancing the confidence level. We then compared this minimum  to that expected for
transit free data. In practice, we simulated 100 light curves without transits using solar data both at high and low activity levels.
We filter the data, search for a periodic signal transit, and calculate the $\chi^2_\text{ratio}$. It turns out that 68\% of the light 
curves will provide a $\chi^2_\text{ratio}>0.971$, while 90\% will be $>0.989$ and 99\% greater than
$0.997$. Thus a transit is detected at a confidence level of 90\% if the $\chi^2_\text{ratio}>0.989$.
These thresholds do not depend on the activity level of the star since they have been pre-filtered.

\subsection{Estimate of the planetary radius}
\label{sec:confid}
\begin{figure}
  \resizebox{\hsize}{!}{\includegraphics{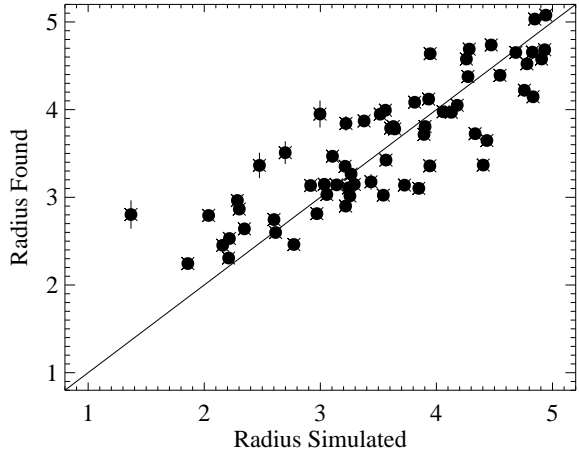}}
 \caption{Correlation between the simulated planet's radius and the one determined by the
Gaussian fit, when the right period has been recovered.}
 \label{fig:correl}
\end{figure}

Since we kept the radius of the parent star equal to 1\,R$_\odot$, the depth of the transit will directly 
estimate of the planet radius. This depth is given by the amplitude of the fitted Gaussian function. In 
Figure~\ref{fig:correl} we show the correlation between the simulated planet's radius and the one determined by the
Gaussian fit, when the right period has been recovered. The maximal deviation between the simulated and detected
radius is  $\sim1$R$_\oplus$.

We also tried to get an estimate of the transit duration, from the width of the fitted Gaussian function; but
due to the high level of noise, no correlation was found for these two parameters.  However, in a second step once the 
transits have been detected, a deeper analysis of the source allows a more accurate fit. An estimation of the transit 
duration is then provided by the width (2$\times$4\,sigma) of the Gaussian function.

\subsection{How the method deals with unevenly sampled data or with short/long data gaps}
The detection method described above has the advantage of not being too affected by unevenly distributed 
data.  The prior filtering of the light curve, on 
the other hand, works properly only for evenly sampled data. This problem can be solved easily by rebinning 
the uneven data set on to a regularly-spaced grid. Concerning the gaps, if short ones are
present (short with respect to the transit duration), this will again affect the results only in the
pre-filter part. A solution will then consist in interpolating the data over the missing values. If medium-sized gaps are
present, the procedure and results will stay the same as long as no transits are present in the gaps. If transits are
present in these gaps, the correct period will not be found even if the dips are very pronounced. When long-lasting gaps are
present, each data set should then be pre-filtered separately. The detection method works also well with long
gaps, depending on their length and the orbital period to be detected. This is of course important if we want to search for planets with periods longer 
than $\approx$~1/3 of the length of the uninterrupted data, and thus have to return to observing the same 
object after an interruption that could be  up to the same length as the original data set (e.g. because 
of reasons of orbital mechanics with respect to the spacecraft). An application of this case is given in
Section~\ref{sec:gaps}.


\section{Performances of the detection method}
\label{sec:results}
\begin{figure}
  \resizebox{\hsize}{!}{\includegraphics{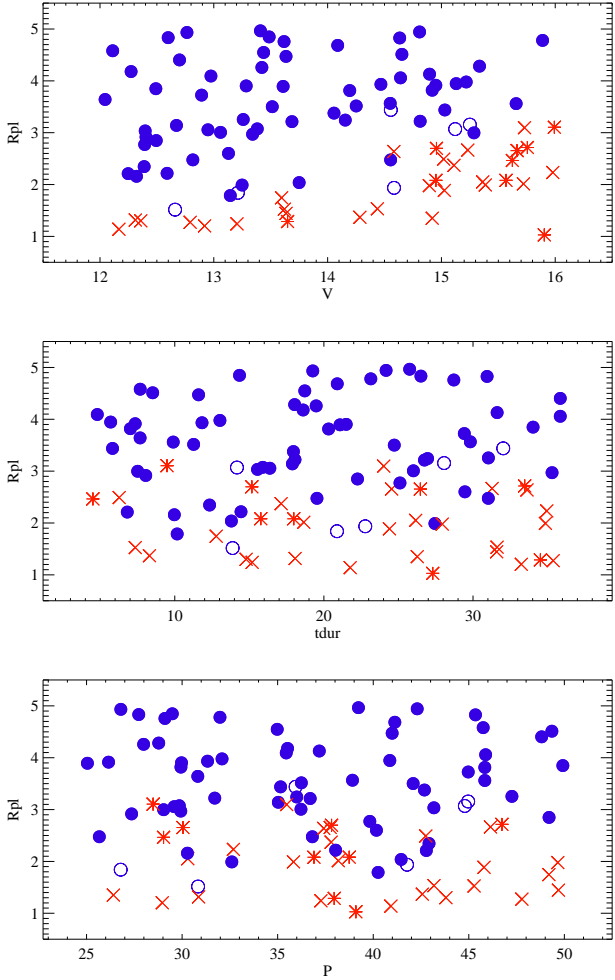}}
 \caption{Results of searching for small planets in 100 lightcurves. Performance of the periodic transit detection algorithm in case of high stellar activity. 
 Dots are associated with a successful recovery of the period with a confidence level $>90\%$ and circles 
 with a confidence level $<90\%$. A failed recovery of the simulated period is shown by a cross 
 when the confidence level is $<90\%$ and by a star for a level $>90\%$. Parameters and
units are explained in Section~\ref{sec:simul}.}
 \label{fig:results}
\end{figure}

\begin{figure}
  \resizebox{\hsize}{!}{\includegraphics{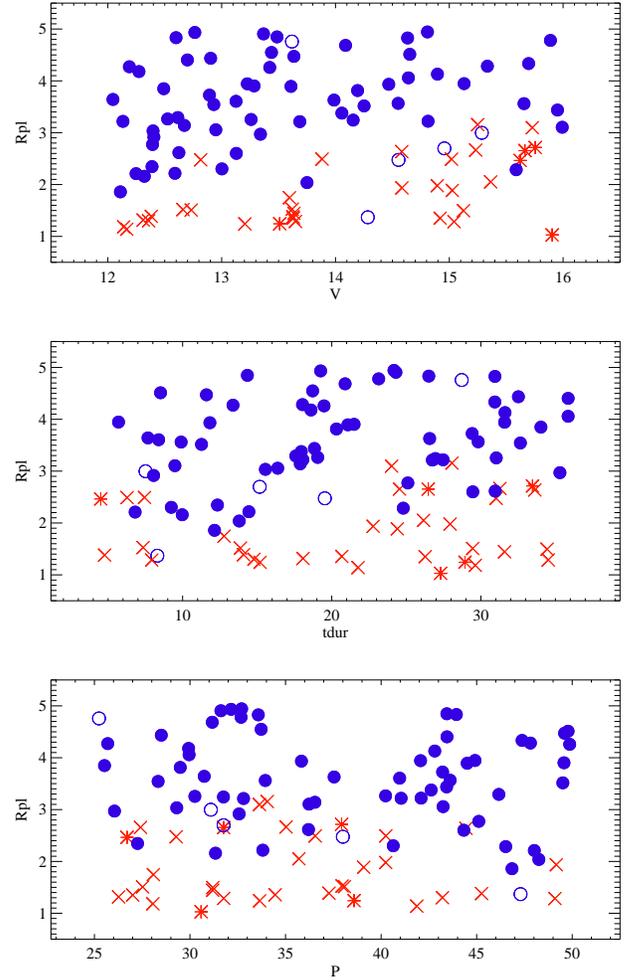}}
 \caption{Same as Figure~\ref{fig:results} but for low stellar activity. }
 \label{fig:results2}
\end{figure}

To test the efficiency of the detection method, we simulated 100 light curves as described in
Section~\ref{sec:simul} (Table~\ref{tab:param}).
Results are displayed in Figure~\ref{fig:results} for the case of high stellar activity and in Figure~\ref{fig:results2} when stellar activity is low. 
We compare the minimum size of the detected planet as a function of the apparent magnitude
(top), the transit duration (middle), and the orbital period (bottom).  Table~\ref{tab:results_high}
shows the total number of simulations (NB., 1st column) for a given range of the input parameters of Table~\ref{tab:param}: radius of the planet,
magnitude, transit duration, and period (related to the number of transits), for high stellar activity level.
The 2nd, 3rd, 4th, and 5th columns show, respectively, the number (in \%) of correct period detection (YES), of failed period
detection (NO), of false alarm (F.A.), and of missed detection (M.D.). 
Table~\ref{tab:results_low} shows the same results but for a low activity level.

\begin{table}
 \centering
 \caption{Performance of the periodic transit detection algorithm in case of high stellar activity.} 
  \label{tab:results_high}
 Planet radius (R$_\odot$)
 \begin{tabular}{cccccc}
 \hline  
 Radius & NB. & YES (\%) & NO (\%) & F.A. (\%) & M.D. (\%)\\
 \hline  
 1--2 & 22& 23& 77& 9& 13\\
 2--3 & 27& 48& 52& 22& 0\\
 3--4 & 31& 94& 6& 3& 10\\
 4--5 & 20& 100& 0& 0& 0\\
  \hline  
   \end{tabular}
   
 Magnitude (Mag)
 \begin{tabular}{cccccc}
 \hline  
 Mag. & NB. & YES (\%) & NO (\%) & F.A. (\%) & M.D. (\%)\\
 \hline  
 12--13 & 26& 81& 19& 0& 4\\
 13--14 & 24& 79& 21& 4& 4\\
 14--15 & 25& 72& 28& 8& 8\\
 15--16 & 25& 36& 64& 24& 8\\
  \hline  
   \end{tabular}
   
 Transit duration (hours)
 \begin{tabular}{cccccc}
 \hline  
 Tdur & NB. & YES (\%) & NO (\%) & F.A. (\%) & M.D. (\%)\\
 \hline  
 4--12 & 22& 77& 23& 9& 0\\
 12--20 & 28& 68& 32& 11& 7\\
 20--28 & 26& 65& 35& 8& 8\\
 28--36 & 24& 58& 42& 8& 8\\
  \hline  
   \end{tabular}
   
 Period (days)
 \begin{tabular}{cccccc}
 \hline  
 P & NB. & YES (\%) & NO (\%) & F.A. (\%) & M.D. (\%)\\
 \hline  
 24--30.5 & 28&  75& 25& 11& 7\\
 30.5--37 & 26&  62& 38& 12& 4\\
 37--43.5 & 25&  68& 32& 8& 4\\
 43.5--50 & 21&  62& 38& 5& 10\\
  \hline  \end{tabular}
\end{table}

\begin{table}
 \centering
 \caption{Same as Table~\ref{tab:results_high} for low stellar activity level.} 
  \label{tab:results_low}
 Planet radius (R$_\odot$)
 \begin{tabular}{cccccc}
 \hline  
 Radius & NB. & YES (\%) & NO (\%) & F.A. (\%) & M.D. (\%)\\
 \hline  
 1--2 & 24& 8& 92& 8& 4\\
 2--3 & 25& 60& 40& 12& 12\\
 3--4 & 30& 93& 7& 0& 0\\
 4--5 & 21& 100& 0& 0& 5\\
  \hline  
   \end{tabular}
   
 Magnitude (Mag)
 \begin{tabular}{cccccc}
 \hline  
 Mag. & NB. & YES (\%) & NO (\%) & F.A. (\%) & M.D. (\%)\\
 \hline  
 12--13 & 32& 75& 25& 0& 0\\
 13--14 & 26& 65& 35& 4& 4\\
 14--15 & 20& 80& 20& 0& 15\\
 15--16 & 22& 41& 59& 18& 5\\
  \hline  
   \end{tabular}
   
 Transit duration (hours)
 \begin{tabular}{cccccc}
 \hline  
 Tdur & NB. & YES (\%) & NO (\%) & F.A. (\%) & M.D. (\%)\\
 \hline  
 4--12 & 21& 71& 29& 5& 10\\
 12--20 & 27& 78& 22& 0& 7\\
 20--28 & 25& 56& 44& 8& 0\\
 28--36 & 27& 59& 41& 7& 4\\
  \hline  
   \end{tabular}
  
 Period (days)
 \begin{tabular}{cccccc}
 \hline  
 P & NB. & YES (\%) & NO (\%) & F.A. (\%) & M.D. (\%)\\
 \hline  
 24--30.5 & 27& 59& 41& 7& 7\\
 30.5--37 & 30& 60& 40& 7& 7\\
 37--43.5 & 24& 67& 33& 4& 0\\
 43.5--50 & 19& 84& 16& 0& 5\\
  \hline  
   \end{tabular}
\end{table}

The main results are that, when assuming solar variability and Poisson noise level as given in Section~\ref{sec:simul}, 
planets with a radius down to $\sim$2\,R$_\oplus$ can be recovered when the parent star has a radius of 1 R$_\odot$. 
The performance depends slightly on the apparent magnitude but not on the transit duration. Finally,
as expected, the efficiency does not depend on the orbital period, which is directly correlated to the
number of transits.


\section{Multiple observations with long and regular gaps}
\label{sec:gaps}
In this section, we investigate one possible application for the \corot mission: what if a planet has an orbital
period close to 1\,yr, when a transit is observed in the first 150\,d data set. Since the satellite cannot
observe the same field again before another 150\,d, would it still be possible to recover the orbital period of the
planet? Since the detection method finds the right period only when all visible transits are
overlayed, the answer is yes.
Again, as in the case of a single observation, a requirement for the period detection
is that the minimum of the global light curve is associated with a transit.

\begin{figure}
  \resizebox{\hsize}{!}{\includegraphics{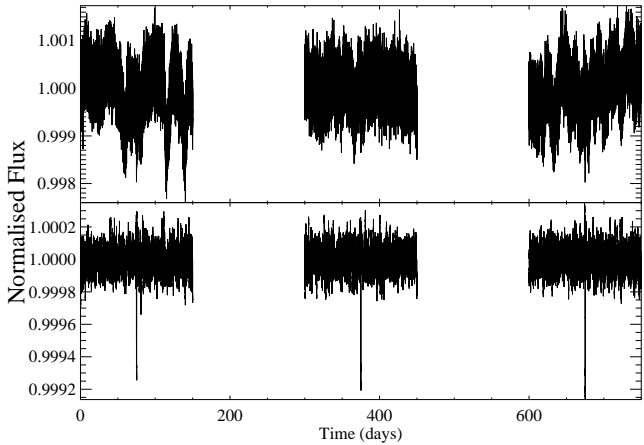}}
 \caption{Same as Figure~\ref{fig:filter} but in the case of three 150\,d observations separated by gaps
 of 150\,d. The phase of the first transit is at the half of the first data set.}
 \label{fig:filter_gaps}
\end{figure}

We simulated three 150\,d observations separated by gaps of the same length. The global light curve then lasts for
about 2.5 yr (see Figure~\ref{fig:filter_gaps} before and after filtering). To ensure that at least one transit is
visible in each data set, we set the first one at the half of the first observation and
allowed the period to change over a low value $P=$300\,d$\pm$37.5\,d. In this exercice we only focused on recover 
periods close to 150\,d.

\begin{figure}
  \resizebox{\hsize}{!}{\includegraphics{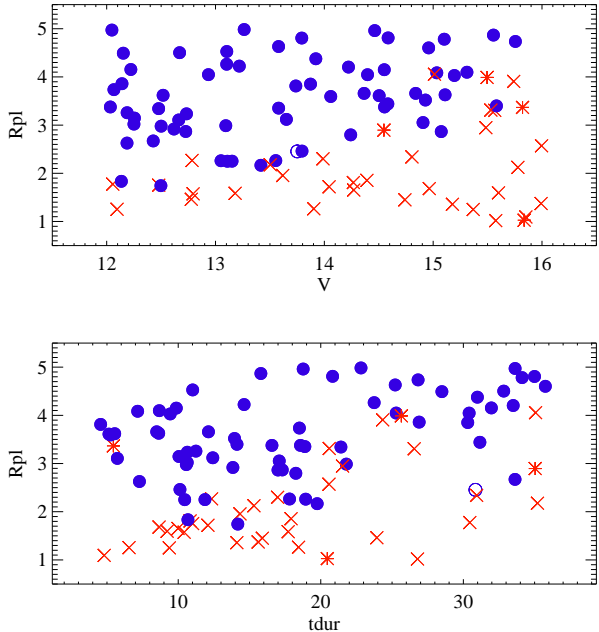}}
 \caption{Same as Figs~\ref{fig:results} and \ref{fig:results2} but in the case of three 150\,d observations separated by gaps
 of 150\,d.}
 \label{fig:results_gaps}
\end{figure}

Figure~\ref{fig:results_gaps} and Table~\ref{tab:results_gaps}, show the same results as Figure~\ref{fig:results} 
and Table~\ref{tab:results_high}, respectively, but for multiple observations. 
The main conclusion from 
Figure~\ref{fig:results_gaps} is that regular longlasting gaps do not  affect the recovery of the transit period when 
close to 150\,d. Results
are similar to the case of one single data set with low stellar activity.
The threshold for the minimum detected planet radius is again around 2\,R$_\oplus$.


\begin{table}
 \centering
 \caption{Same as Table~\ref{tab:results_high} but in the case of three 150\,d observations separated by gaps
 of 150\,d.} 
  \label{tab:results_gaps}
 Planet radius (R$_\odot$)
 \begin{tabular}{cccccc}
 \hline  
 Radius & NB. & YES (\%) & NO (\%) & F.A. (\%) & M.D. (\%)\\
 \hline  
 1--2 & 23& 9& 91& 4& 0\\
 2--3 & 23& 65& 35& 4& 4\\
 3--4 & 29& 83& 17& 7& 0\\
 4--5 & 25& 96& 4& 0& 0\\
  \hline  
   \end{tabular}
   
 Magnitude (Mag)
 \begin{tabular}{cccccc}
 \hline  
 Mag. & NB. & YES (\%) & NO (\%) & F.A. (\%) & M.D. (\%)\\
 \hline  
 12--13 & 28& 79& 21& 0& 0\\
 13--14 & 24& 79& 21& 0& 4\\
 14--15 & 23& 65& 35& 4& 0\\
 15--16 & 25& 36& 64& 12& 0\\
  \hline  
   \end{tabular}
   
 Transit duration (hours)
 \begin{tabular}{cccccc}
 \hline  
 Tdur & NB. & YES (\%) & NO (\%) & F.A. (\%) & M.D. (\%)\\
 \hline  
 4--12 & 32& 69& 31& 3& 0\\
 12--20 & 31& 65& 35& 0& 0\\
 20--28 & 18& 50& 50& 11& 0\\
 28--36 & 19& 74& 26& 5& 5\\
  \hline  
   \end{tabular}
  
\end{table}


\section{Summary}
\label{sec:conc}
In this paper, we have presented a new method for detecting transits of a planet orbiting around its parent star, and we
investigate the limit of the smallest detectable planet when assuming that the star has similar characteristics to the
Sun. We simulated light curves as described in Section~\ref{sec:simul}, and, before applying the transit detection algorithm, 
we prefiltered the data to removing low and high frequency noise.

Assuming a single observation of 150\,d, the period of at least 3 transits from a planet with a radius down to 
$\sim$2\,R$_\oplus$ 
can be recovered. The detection method does slightly depend on the apparent magnitude, but not on 
the transit duration in the given range of parameter values. The recovery of the right period
does not depend on the number of transits, which only affects the confidence level (more transits will increase the confidence level).

We have also finally investigated the case of multiple observations (3$\times$150\,d) of the same field, when regular gaps of
150\,d are present and when one single transit is visible in all data sets. In this case, assuming the period is close to 150\,d, the limit for the
smallest detected planet  does not differ from the case when all transits are observed in a single data set.
These multiple observations may represent an application for the \corot or similar space missions to recover planetary transits when the orbital period is 
much longer than the single observation duration.

\begin{acknowledgements}
The availability of the VIRGO/SoHO data on total solar irradiance and spectral irradiances from the VIRGO Team through PMOD/WRC (Davos, Switzerland) 
and of unpublished data from the VIRGO Experiment on board  the ESA/NASA Mission SoHO are gratefully acknowledged. 
We also gratefully acknowledge very constructive and instructive comments by S. Aigrain and by H. Deeg.
\end{acknowledgements}

\end{document}